\documentstyle[epsf]{l-aa}

\def\la{\mathrel{\hbox{\rlap{\hbox{\lower4pt\hbox{$\sim$}}}\hbox{$<$}}}}
\def\ga{\mathrel{\hbox{\rlap{\hbox{\lower4pt\hbox{$\sim$}}}\hbox{$>$}}}}

\def\arcmin{\hbox{$^\prime$}}
\def\arcsec{\hbox{$^{\prime\prime}$}}

\def\fm{\hbox{$.\!\!^{m}$}}
\def\fs{\hbox{$.\!\!^{s}$}}

\def\micron{\hbox{$\mu$m}}

\newcommand{\etal}{{\it et al.}\,}      
\newcommand{\ie}{{\it i.e.},\ }         
\newcommand{\cf}{{\it cf.},\ }          
\def\deg{{^\circ}}

\newcommand{\kms}{{\,km\ s$^{-1}$\,}}

\newcommand{\HI}{\mbox{\normalsize H\thinspace\footnotesize I}}

\newcommand{\DVT}{$\Delta v_{20}$}
\newcommand{\DVF}{$\Delta v_{50}$}

\begin{document}
\thesaurus{03(09.04.1; 
              11.03.4; 
              11.04.1; 
              11.09.1; 
              11.16.1; 
              11.19.1) 
                      }

\title{Multiwavelength Observations of a Seyfert 1 Galaxy Detected in ACO 3627.}
 
\author{P.A.~Woudt\inst{1,5} \and R.C.~Kraan-Korteweg\inst{2,6} 
\and A.P.~Fairall\inst{1}  
\and H.~B\"ohringer\inst{3} \and
V.~Cayatte\inst{2} \and I.S.~Glass\inst{4}}
              
\offprints{Patrick A. Woudt}
 
\institute{Department of Astronomy, University of Cape Town, 
Rondebosch, 7700 South Africa
\and
Observatoire de Paris-Meudon, D.A.E.C., 5 Place Jules Janssen,
92195 Meudon Cedex, France
\and
Max-Planck-Institut f\"ur extraterrestrische Physik, D-85740 Garching, Germany
\and
South African Astronomical Observatory, P.O.~Box 9, Observatory 7935, 
Cape Town, South Africa
\and
European Southern Observatory, Karl-Schwarzschildstr.~2, D-85748 Garching, Germany
\and
Departemento de Astronomia, Universidad de Guanajuato, Apartado Postal 144, 
Guanajuato, Gto 36000, Mexico}
 
\date{Received date; Accepted date}

\maketitle

\markboth{P.A.~Woudt et~al.: A Seyfert 1 Galaxy in ACO 3627}
{Woudt et al.}

\begin{abstract}

ACO 3627 is a rich, nearby cluster of galaxies at the core
of the Great Attractor. At the low galactic latitude of 
$b = -7.2\deg$ the galactic extinction is significant. Nevertheless, 
its proximity makes it a prime target for studies of environmental 
effects on its cluster members. Here, we report on a  
multi-wavelength study of a Seyfert 1 galaxy at 30 arcmin
from the centre of ACO 3627. Its Seyfert nature was discovered 
spectroscopically and confirmed in X-rays. We have obtained 
B$_{\rm J}$ and  R$_{\rm C}$ CCD photometry as well as J, H, K and 
L aperture photometry at the SAAO, low and high resolution spectroscopy 
(ESO and SAAO), 21 cm line observations (Parkes Observatory) and X-ray ROSAT 
PSPC data. 

The Seyfert 1 galaxy is of morphology SBa(r). It has a nearby companion
(dS0) but shows no signs of interaction. A consistent value for the galactic 
extinction of A$_{\rm B}$ = 1.6 mag could be determined. The nucleus of the 
Seyfert is very blue with a strong (B$_{\rm J}$ -- R$_{\rm C}$) colour gradient 
in the inner 2.5 arcsec. The extinction-corrected near-infrared 
colours of WKK 6092 are typical of a Seyfert 1 and the X-ray spectrum 
conforms to the expectation of a Seyfert as well. The galaxy has a very 
low \HI\ flux. This could be explained by its morphology, but also -- 
due to its very central position within the rich Norma cluster --  
to ram pressure stripping.

\end{abstract}


\section{Introduction}

Dust and stars in the Milky Way obscure a large fraction of the extragalactic
sky, creating a ``Zone of Avoidance'' (ZOA) in the distribution
of galaxies. 
In an effort to reduce the size of the ZOA and thus coming closer
towards an all-sky distribution of galaxies, we have embarked on a deep optical
galaxy search behind the southern Milky Way (Kraan-Korteweg \& Woudt 1994). 
This has led to the recognition that ACO 3627 (Abell \etal\ 1989), also
called the Norma cluster after the constellation it is located in, is a 
massive, nearby cluster of galaxies at the core of the Great Attractor (GA) 
$(\ell,b,v) = (325\deg, -7\deg, 4882$ \kms) (Kraan-Korteweg \etal\ 
1996).
The Norma cluster appears to be the central, dominant component of a 
``great wall''-like structure and would be the most prominent overdensity
of galaxies in the southern sky, were it not obscured by the Milky Way 
(Woudt \etal\ 1997).

Recent observations of the Norma cluster with the ROSAT PSPC 
have confirmed the massive nature of ACO 3627; it is the 
6$^{th}$ brightest cluster in the ROSAT sky (B\"ohringer \etal\ 1996). The
X-ray contours furthermore suggest the existence of a subcluster.
The merging scenario is independently supported by the radio continuum 
emission of the central cD galaxy PKS1610-608. The emission from
this wide-angle-tail (WAT) radio source  (Jones \& McAdam 1992) seems 
to encircle the X-ray subcluster (\cf\ Fig.~3 of Kraan-Korteweg
\etal\ 1997) and is indicative of a strong motion of the cluster
gas due to the ongoing merging process 
(Jones \& McAdam 1996, Burns \etal\ 1994).

Roughly 30{\arcmin} from the centre of this cluster -- taken as the
central cD galaxy PKS1610-608 -- we have identified a Seyfert 1 galaxy.
It is a member of ACO 3627. In the following sections we describe the 
various observations of this galaxy: 
the discovery of the galaxy in section 2,
the multicolour photometry obtained at the South African Astronomical 
Observatory (SAAO) in section 3, 
the spectroscopy obtained at the European Southern Observatories (ESO) 
and the SAAO in section 4, 
the \HI\ observations obtained with the 64m radio telescope of the 
Parkes Observatory of the Australian Telescope National Facility (ATNF)
in section 5, 
and the X-ray data from ROSAT PSPC observations in section 6. 
The results are summarized and discussed in the last section.

\section{Discovery of the Seyfert 1}

Two of us (RKK and PAW) have been engaged in a deep optical
search for galaxies behind the southern Milky Way ($265\deg \la \ell \la
340\deg, |b| \la 10\deg$) (Kraan-Korteweg \& Woudt 1994). 
With a 50x magnification, film copies of the SRC IIIaJ survey were 
systematically scanned by eye and all galaxies with $D \ge 0.2\arcmin$ 
were catalogued. So far, over 11000 previously unknown galaxies have been 
identified in this region of the sky. 

The galaxy that we report on here, WKK 6092 (Woudt \& Kraan-Korteweg 1998) 
has been classified by us as a barred, early type spiral 
SBa. Its observed dimensions are D x d = 56{\arcsec} $\times$ 47{\arcsec} 
with an estimated blue magnitude of B$_{\rm J}$ = 14\fm7. It had not been 
catalogued before by Lauberts (1982) nor is it listed in the IRAS Point 
Source Catalogue. 

\begin{figure}
  \epsfxsize 8.4cm \epsfbox{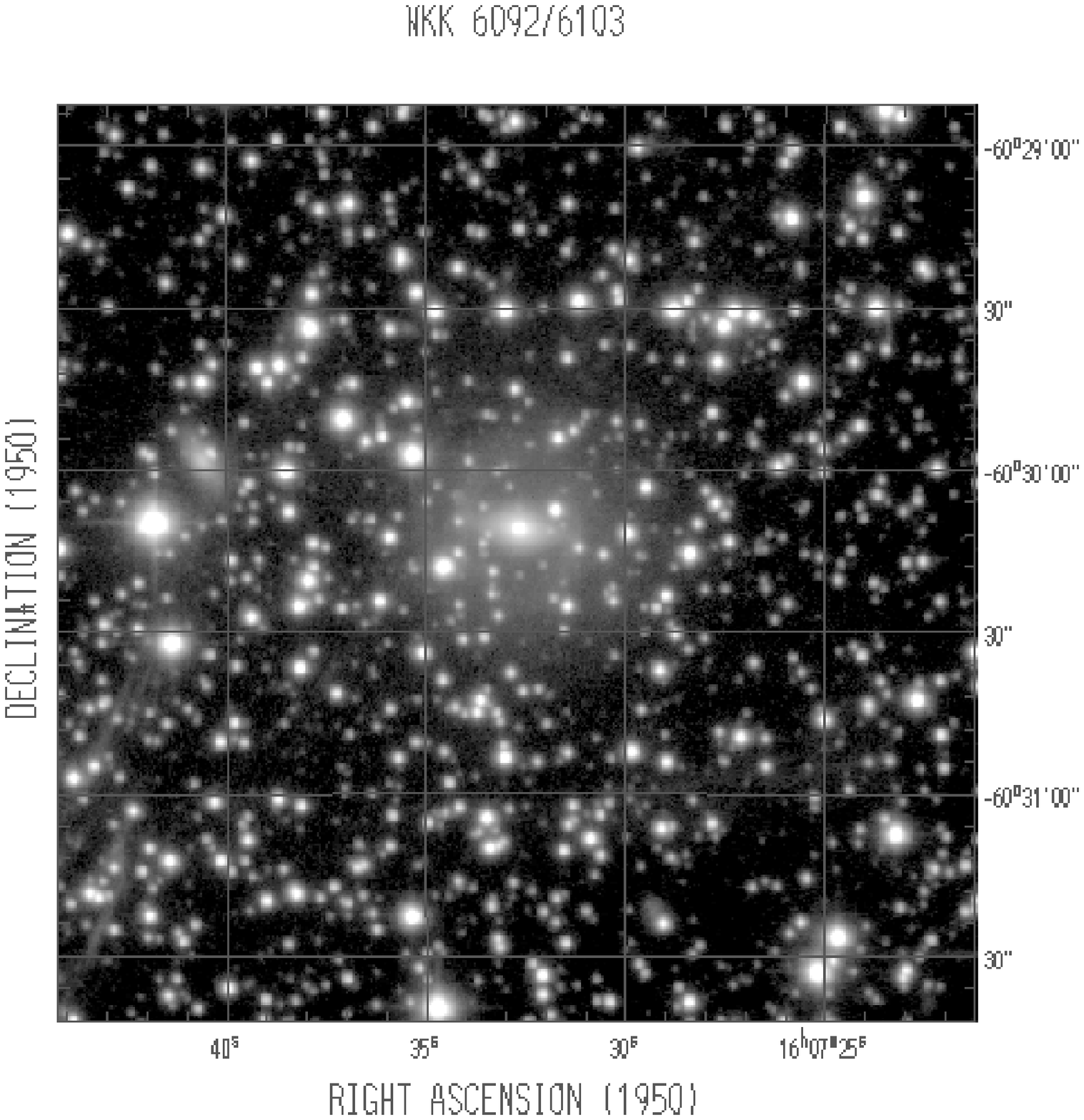}
  \vspace{-0.02cm}
  \hspace{0.5cm}
  \epsfxsize 7.0cm \epsfbox{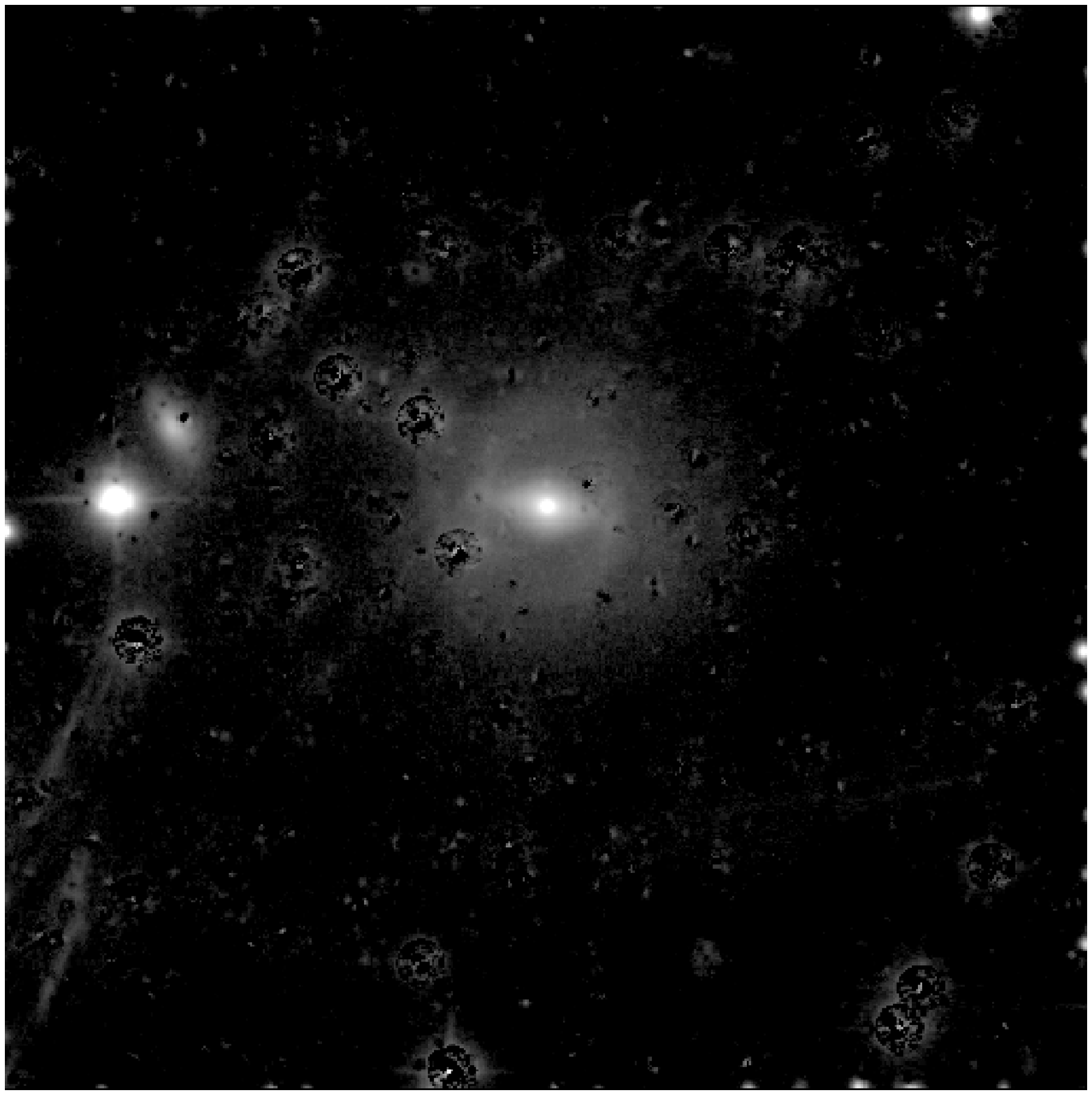}
  \caption{The R$_C$ CCD image of WKK 6092. Right Ascension and Declination 
   (1950) are labelled. The companion WKK 6103 (R.A. = 16$^h$07$^m$40\fs6,
   Dec = -60$^\circ$29$^m$58$^s$) is  to the east of WKK 6092.
   The lower panel shows the same image with the stars substracted.}
  \label{ccdim}
\end{figure}

Approximately 1 arcmin to the east of WKK 6092, another galaxy, WKK 6103,
was found. This galaxy is an early type spiral of 
dimensions 19{\arcsec} $\times$ 9{\arcsec} and B$_{\rm J}$ = 17\fm1. Both 
galaxies are displayed in Fig.~\ref{ccdim}.

\section{Spectroscopy}

The spectroscopic observations described here are part of a larger
program to gauge the 3-dimensional distribution of galaxies behind the
southern Milky Way. The reader is referred to Kraan-Korteweg \etal\ 
(1994) for details and preliminary results on the 
individual approaches.

\subsection{MEFOS spectroscopy}

The MEFOS ({\bf M}eudon {\bf E}SO {\bf F}ibre {\bf O}bject {\bf S}pectrograph) 
observation of WKK 6092 was made in February 1994. 
MEFOS is mounted at the prime focus of the 3.6-m 
telescope of the European Southern Observatory (ESO), La Silla, resulting 
in a 1 degree field (Felenbok \etal\ 1997).
The CCD Tek \#32 was used together with grating
\#15, yielding a dispersion of 170\AA/mm and a resolution of about 11\AA.
The exposure time was 2 x 30 minutes.

\begin{figure}
  \epsfxsize 8.4cm \epsfbox{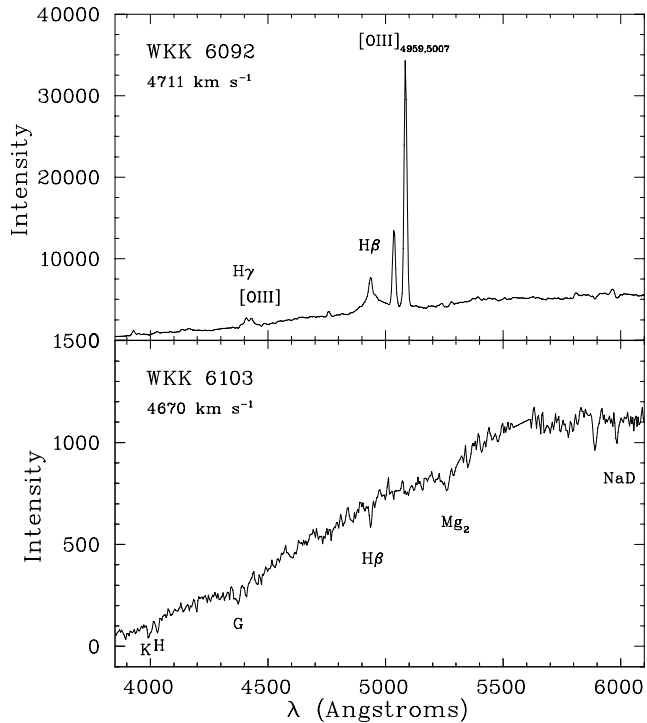}
  \caption{The MEFOS spectrum of the Seyfert WKK 6092 (upper panel) and 
the companion WKK 6103 (lower panel).}
  \label{mefspec}
\end{figure}

The MEFOS spectrum of WKK 6092 (\cf\ Fig.~\ref{mefspec}, upper panel) revealed 
broad Balmer emission 
lines, indicative of it being a Seyfert 1 galaxy. The emission lines yield 
an observed radial velocity of $v_{em} = 4711 \pm 30$ {\kms}. Further 
analysis is not possible from this low resolution, narrow wavelength range 
(3850\AA -- 6100\AA) spectrum, and the galaxy was subsequently reobserved 
at the SAAO (\cf\ Fig.~\ref{saaospec}, next section). The neighbouring galaxy WKK 6103
-- visible to the left of the Seyfert in Fig.~\ref{ccdim} -- has been observed during
the same run on another MEFOS field. It has a radial velocity of 
$v_{abs} = 4670 \pm 45$ \kms (determined from standard cross-correlation 
techniques following Tonry and Davies 1979), is hence at a similar redshift 
as the Seyfert and might be a companion of WKK 6092. 

Both spectra are displayed in Fig.~\ref{mefspec}. The upper panel shows the Seyfert and
the lower panel the neighbouring galaxy.

\subsection{S.A.A.O. spectroscopy}

We reobserved WKK 6092 for a total of 2500 seconds in June 1995 at the SAAO 
using the 1.9-m Radcliffe reflector with ``Unit'' spectrograph and a reticon 
photon-counting detector. This results in a wider spectral range 
(3500\AA -- 7500\AA) including therewith also the 
H$\alpha$ emission line. More importantly, we observed a
spectro-photometric standard star (LTT 7379) allowing the determination
of the relative instrumental response and its correction.

\begin{figure}
  \epsfxsize 8.4cm \epsfbox{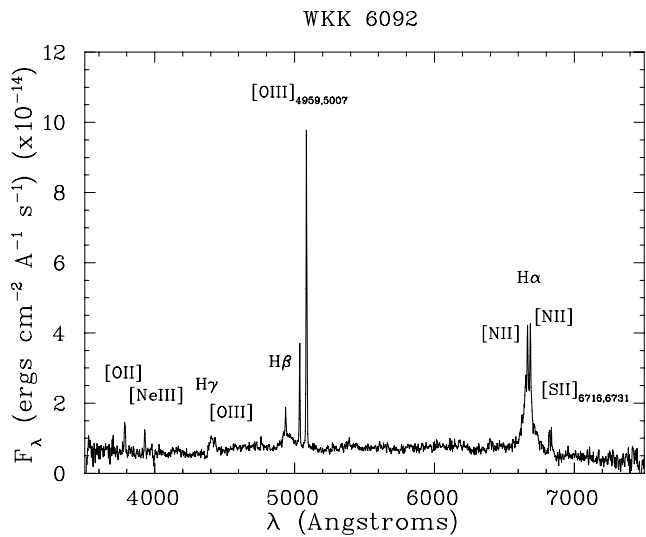}
  \caption{Flux calibrated SAAO spectrum of the Seyfert 1 WKK 6092. 
  The spectrum was calibrated using the spectrophotometric standard star
  LTT 7379. The emission lines are labelled.}
  \label{saaospec}
\end{figure}

The spectrum is displayed in Fig.~\ref{saaospec}. Its features identify it as a Seyfert 1 
galaxy: broadened Balmer lines with a full width at zero intensity of $7500$ 
{\kms}, but with significant narrow emission superposed, and strong narrow 
forbidden lines (\cf\ labels in Fig.~\ref{saaospec}). 

The line strengths could be measured to an accuracy of approximately 10{\%}. For the 
narrow emission lines, the uncertainty is somewhat higher due to
the difficulty in separating the broad from the narrow emission lines, particularly in H$\alpha$ 
where the peak of the broad component is heavily masked by the [NII] lines on either side of the narrow
H$\alpha$ emission line. The line strengths are listed in Table 1. The header line lists the identified 
emission lines with their respective wavelengths in {\AA}ngstr{\o}m.
The fluxes are expressed relative to a total H$\beta$-line intensity of 100.

The broad emission lines show a Balmer decrement (H$\alpha$/H$\beta$) of 4.24, the 
narrow lines indicate a general Balmer decrement of 4.47. Given the difficulty in separating the
broad from the narrow component in the SAAO spectrum, these values are identical
within the observational errors.

The steepening in the Balmer decrement of the narrow component -- the recombination
value is 2.85 (Aller 1984) -- is caused by reddening alone
(Osterbrock 1989) and corresponds to a galactic extinction of A$_{\rm B} = 1.7$ mag.
This is in good agreement with values derived from the galactic
\HI\ column density which -- at the position of the Seyfert -- is 
N$_{\rm HI} = 2.05 \cdot 10^{21}$atoms cm$^{-2}$ (Kerr \etal\ 1986). Assuming
a constant gas-to-dust ratio, the formalism given by Burstein \& Heiles (1982)
predicts an absorption in the blue of A$_{\rm B}$=1.5 mag. This value is furthermore
supported by extinction measurements from fits to the ROSAT PSPC X-ray 
spectrum (\cf section 6).

\begin{table*}
\normalsize
\caption{Line strengths determined from the flux calibrated SAAO spectrum.}
\begin{tabular*}{17.5cm}{l @{\extracolsep\fill}*{9}{c}}
\noalign{\smallskip} 
\hline
\hline
\noalign{\smallskip} 
               & H$\beta$ Broad & H$\beta$ Narrow & [OIII] & [OIII] & H$\alpha$ Broad & H$\alpha$ Narrow & [NII] & [SII] & [SII] \\
               & 4861{\AA}     & 4861{\AA}     & 4959{\AA}   & 5007{\AA}   & 6563{\AA}      & 6563{\AA}      & 6584{\AA}  & 6716{\AA}  & 6731{\AA}  \\
\noalign{\smallskip}
\hline
\noalign{\smallskip}
 WKK 6092    & 85       & 15       & 52     & 160    & 360       & 67        & 63    & 14    & 13    \\
\noalign{\smallskip}
\hline
\hline
\end{tabular*}
\end{table*}

The ratio of the [SII] lines 6716/6731 indicates an electron density (N$_e$)
of approximately 600 cm$^{-3}$. Generally high excitation is shown by the 
presence of weak [OIII] 4363 and [NeIII] 3869. The strength of the 4363 line
suggests a temperature above 15000 K. The continuum can be described 
by a power law.

\section{Photometry}
\subsection{B$_{J}$ and R$_{C}$ CCD surface photometry}

In April 1996, we obtained B$_{\rm J}$ and R$_{\rm C}$ CCD images of WKK 6092 with
the 1-m telescope of the SAAO. The TEK \#8 CCD was used, giving a field 
of view of 2.6{\arcmin} $\times$ 2.8{\arcmin} and a pixel size of 
0.347 arcsec/pixel.

WKK 6092 was observed under good seeing conditions (1.2{\arcsec} and 
1.3{\arcsec} for R$_{\rm C}$ and B$_{\rm J}$ respectively) with exposure times of
1200 seconds in R$_{\rm C}$ and 2400 seconds in B$_{\rm J}$. 
The R-band image is displayed in the upper panel of Fig.~\ref{ccdim}.
This image clearly illustrates the inherent difficulties in identifying 
galaxies at these low galactic latitudes. Two of the more serious 
complications are:

\begin{enumerate}
 \item{The overwhelming number of foreground stars; applying DAOPHOT
a thousand stars were, for instance, detected 3$\sigma$ above the
background on the R-band image.}
 \item{The uncertain and non-uniform galactic foreground extinction 
at these low galactic latitudes ($|b| \le 10^{\circ}$) and its effect on the
observed properties of the regarded galaxies.}
\end{enumerate}

In the lower panel of Fig.~\ref{ccdim}, the star-substracted CCD image is displayed.
This image demonstrates the effectiveness of the star substraction
routine (important for the determination of the magnitude) and 
reveals further detailed structure of the Seyfert.
The galaxy has a very blue nucleus. The bar of the galaxy is 
quite distinct and the disk very smooth with a clear superimposed 
ring; it is in fact a SBa(r) galaxy. The 
features of the Seyfert and the nearby companion, a dS0, 
do not reveal any indication that this galaxy pair is in gravitational 
interaction, despite their close position on the sky and in 
velocity space.

To determine the radial surface brightness profile and total apparent 
magnitudes one must first detect and then mask all the foreground stars in the
image. This was done with DAOPHOT as implemented in IRAF. Once all the stars
are identified, stellar light above a certain isophotal treshold is masked, 
this is approximately 23 mag/${\Box}''$ in R$_{\rm C}$ and 24 mag/${\Box}''$ 
in B$_{\rm J}$.

The surface brightness profile is determined using Jedrzejewski's 
method (Jedrzejewski 1987) in IRAF. 
Both the ellipticity and position angle are kept fixed whilst determining
the radial profile. This way the average counts per pixel will be based on the
unmasked pixels. This method only works if less than 50\% 
of the light is masked within the ellipse.
A more detailed description of the surface photometry will
be given in a separate paper, where we will map the galactic 
extinction from  B$_{\rm J}$ and R$_{\rm C}$ photometry combined with the
Mg$_2$ index of elliptical galaxies (Woudt \etal\ 1998).

\begin{figure}
  \epsfxsize 8.4cm \epsfbox{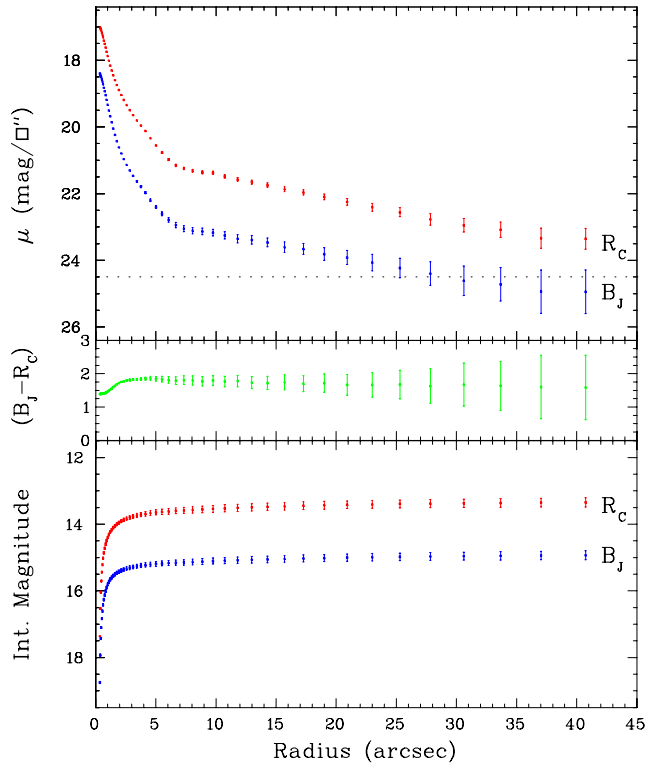}
  \caption{Radial surface brightness profile (upper panel), Radial colour
profile (middle panel) and the integrated magnitudes (lower panel).}
  \label{radprof}
\end{figure}

The upper panel of Fig.~\ref{radprof} shows the radial surface brightness profile
of the Seyfert galaxy. The dotted line at the B = 24.5 mag/${\Box}''$
isophotal level corresponds to the isophotal level of our diameter estimate 
on the IIIaJ film copy (Kraan-Korteweg \etal\ 1995). Our 'eye' estimate
of 56'' x 47'' from the IIIaJ SRC film copy agrees very well with the 
CCD data (D x d)$_{\rm B=24.5}$ = 58'' x 52''.
 
With an observed axial ratio of  d/D = 0.89 and an intrinsic flattening 
of r$_o$ = 0.2, the inclination of the Seyfert according
to the formalism given by Holmberg (1946), cos$^2$i = (r$^2$ 
- r$_o^2$)/(1 - r$_o^2$) is i = 28$\deg$.

The middle panel of Fig.~\ref{radprof} shows the (B$_{\rm J}$ -- R$_{\rm C}$) colour. The inner 2.5 
arcsec reveal a strong gradient: within the inner area the (B$_{\rm J}$ -- R$_{\rm C}$) 
colour changes by 0.5 -- 0.6 mag. Not surprisingly, the nucleus of the Seyfert 
is very blue.

The lower panel shows the integrated magnitude as a function of radius.
Within each ellipse the sum of the masked and unmasked pixels is multiplied
by the average counts per pixel and this is then integrated over the entire
galaxy. The total magnitude at the asymptotic value was found to be
B$_{\rm T}$ = $14.88 \pm 0.13$ mag, respectively R$_{\rm T}$ = $13.30 \pm 0.14$ mag.

\subsection{Near-Infrared observations}

The 0.75-m telescope of SAAO was used to obtain single aperture (9'')
JHKL (1.25 -- 3.4 {\micron}) broadband photometry of WKK 6092.
The Seyfert was observed twice in May 1996. The observations 
were made in the same photometric system as reported by Glass \& 
Moorwood (1985) and further details on observation and reduction
procedures are described there. 

The resulting J, H, K and L magnitudes of WKK 6092 are 12.91, 11.91, 
11.51 and 10.60 (\cf Table~\ref{seytab}), and the respective near infrared colours 
are (J -- H) = 0.93, (H -- K) = 0.47, (K -- L) = 0.91. The typical
errors in J, H and K are 0.03 mag, whereas the error in L is somewhat 
larger (0.15 mag).

Assuming A$_{\rm B}$ = 1.6 mag (\cf\ section 3.2 and 6) and the relative 
extinction values in the near infrared by Cardelli \etal\ (1989), 
the extinction-corrected colours are
(J -- H)$_0$ = 0.82, (H -- K)$_0$ = 0.38, (K -- L)$_0$ = 0.84.
These colours compare well with known Seyfert 1's (e.g. NGC 1566), although
there is evidence for a significant contribution of the underlying galaxy
-- even with the 9'' aperture which primarily contains the nucleus.

\section{21cm observations}

The Seyfert galaxy was observed in \HI\ with the 64-m radio
telescope at the Parkes Observatory (ATNF) in July 1995
as part of program of measuring redshifts for obscured, low 
surface brightness galaxies behind the Milky Way (Kraan-Korteweg 
\etal\ 1998). In this
observing mode we used two IF's and offset 512 of the 1024 channel
of each polarization by 22 MHz. This resulted in a
velocity coverage of $0-10000$ \kms and a velocity resolution 
of $12.6$ \kms. 
The observations were carried out in total power 
mode. Six 10 minute ON-source observations were performed, preceded 
by an equal length OFF-source observation at the same Declination 
but 10.5 time minutes earlier in Right Ascension to cover the 
same patch of the sky for both the reference and the signal spectrum.
The online control program automatically corrected for the zenith angle
dependence of the telescope sensitivity. The telescope has a HPBW of
15$\arcmin$ with a system temperature at 21 cm of typically T$_{sys}$
= 39 Jy.  

\begin{figure}
  \epsfxsize=8.4cm \epsfbox[30 30 560 391]{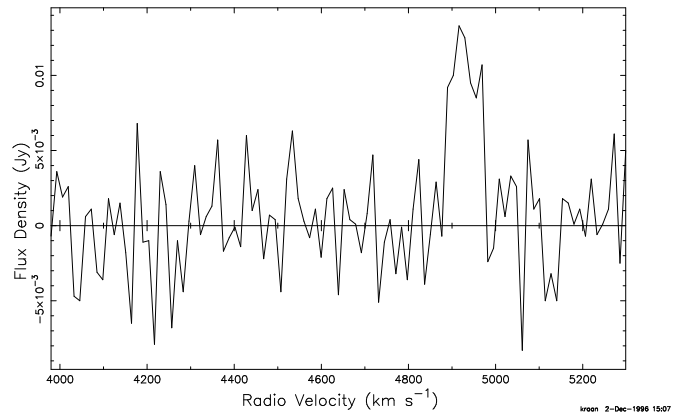}
  \caption{\HI\ spectrum as obtained at the Parkes 64m radio telescope
  from a 60 minute integration. The velocity axis is in the radio convention.}
  \label{hispec}
\end{figure}

The data were reduced using the package ``SLAP'', or Spectral Line 
Analysis Program (Staveley-Smith 1985). The two polarisations
of the 6 scans were averaged together. With a 5th-order polynomial
baseline an r.m.s. of $\sigma_f$ = 3.3 mJy was  obtained for this 60 
minute integration. The reduced \HI\ spectrum of the Seyfert galaxy 
is displayed in Fig.~\ref{hispec}. 

A weak signal is seen at v$_{rad} = 4929$ {\kms}, translating to a
velocity in the optical convention of v$_{opt} = 5012$ {\kms}. This is
higher than the optical measurements ($\Delta$v = 301 
respectively 324\kms for MEFOS and SAAO). 
A difference between the HI velocity and the optical measurement, as observed here
for WKK 6092, has been seen before in Seyfert galaxies (Mirabel \& Wilson 1984). It is caused
by the net outflow of gas in the narrow emission-line regions (Mirabel \& Wilson 1984). The observed
velocity difference for WKK 6092, \ie $\sim$300 {\kms}, is relatively large for Seyfert 1's.

The signal to noise ratio is only S/N = 2.8, however, the signal is consistent 
in flux and shape throughout all individual scans. The profile is 
narrow and Gaussian in shape -- not unexpected considering that 
this galaxy is fairly face-on.

The linewidths measured at the 50\%  and 20\% level are \DVF = 88\kms 
and  \DVT = 97\kms. Even when corrected for inclination and a
velocity dispersion in the z-direction (\cf\ Richter \& Huchtmeier 1984) 
the linewidths (150 and 169 {\kms}) seem relatively low for a luminous 
spiral galaxy. The integrated \HI\ flux of the Seyfert is 
I  = 0.93 Jy {\kms}. 

Adopting a distance of 93 $h_{50}^{-1}$ Mpc (Kraan-Korteweg \etal 1996),
global properties can be evaluated. The \HI\ mass is:
$${\cal M}_{HI} = 2.36 \cdot 10^5 \cdot R^2 \cdot I 
                = 1.9 \cdot 10^9 {\cal M}_{\odot}.$$
The absolute blue magnitude, corrected for galactic foreground extinction, is M$_{B_T}^o = -21.52$. 
The \HI\ mass-to-blue light ratio then is:
$${\cal M}_{HI} / {\cal L}_{B^o} = 0.03.$$ 
Following Casertano \& Shostak (1980), the total mass can be determined 
from the corrected linewidth, the distance R (in Mpc) and
the extinction-corrected diameter D$^o$ (in arcmin):
$${\cal M}_T = 7500 \cdot D^o \cdot R \cdot (\Delta v_{20}^{o,i})^2 
             = 3.0 \cdot 10^{10} {\cal M}_{\odot}.$$ 

Although the \HI\ mass and \HI\ mass-to-blue light ratio are quite
low, they both lie well within the range typical of barred spiral
galaxies of morphological type SBa to SBab (\cf\ Huchtmeier \& Richter 1988, 
1989 for a field and cluster sample). Considering furthermore
the uncertainties due to the large foreground extinction corrections,
the total mass is also in perfect agreement with the expectation for a
barred early-type spiral.

\section{X-ray}

The cluster ACO 3627 was observed with the ROSAT PSPC for an effective
exposure time of 11,257 ksec on September 23 and 24, 1992 and March 
13 to 15, 1993. In this observation WKK 6092 is well detected as
an X-ray source about 6 arcsec to the north of the optical position. 
This is well within the typical pointing accuracy of the ROSAT observatory 
and the Seyfert galaxy can actually be used for the astrometric correction 
of this ROSAT pointing. The Seyfert galaxy is located at a distance of 
about 20 arcmin from the X-ray cluster centre (note: not the central 
radio source). It lies in the outer very low surface brightness
region of the cluster (\cf\ images in B\"ohringer \etal\ 1996) and
stands out clearly above the cluster background 
and the general X-ray background. In the PSPC observation the Seyfert
is located very close to the support ribs of the PSPC window.
With the galaxy approximately in the same direction from the
pointing centre (cluster centre) as the position angle of the wobbling
of the ROSAT telescope, the X-ray source is, however, not so severely 
obscured and a good flux measurement is possible.

About $580 \pm 80$ source counts are detected for the Seyfert, yielding a
count rate of $0.052 \pm 0.007$ s$^{-1}$, an unabsorbed (0.5 - 2.0 keV) flux of
$1.05 \pm 0.15 \cdot 10^{-12}$ erg s$^{-1}$ cm$^{-2}$ and a luminosity  of
$1.2 \pm 0.17 \cdot 10^{42}$ erg s$^{-1}$ in the 0.5 to 2.0 keV band.  
No significant variation of the count rate could be detected within and
between the observing intervals.

\begin{figure}
  \epsfxsize 8.4cm \epsfbox{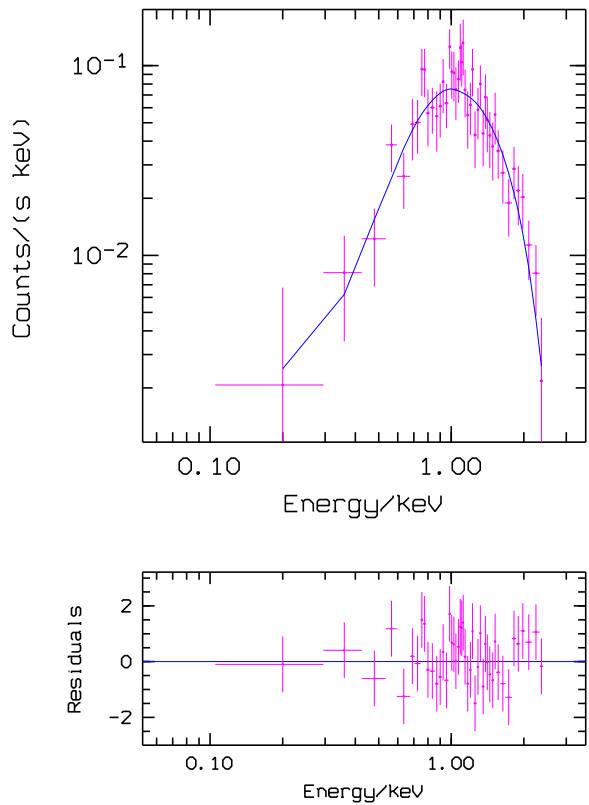}
  \caption{X-ray spectrum WKK 6092.}
  \label{seyx}
\end{figure}

The background substracted spectrum of WKK 6092 is shown in Fig.~\ref{seyx}. It can 
be perfectly fitted with an absorbed power law model with best fitting
values for the photon power law index of about $\gamma = -1.7$ and the
absorption column density of about $N_H \sim 2.2 \cdot10^{21}$ cm$^{-2}$
yielding a reduced $\chi ^2$ of 0.8. But the uncertainties are very large 
with values of the order of 70\% ($1 \sigma$). The photon index is
equal to the canonical value and the absorption column density is close to
the value found for the cluster, and independenly from 21 cm observations
at this position (Kerr \etal\ 1986, \cf\ also section 3.2). We can 
therefore only state that the spectral fits are consistent with the standard
expectation for this object. An upper limit (at a 2 $\sigma$ level) for the internal absorption
of the X-ray emission from the Seyfert galaxy can be set to about
$6 \cdot 10^{21}$ cm $^{-2}$. 

\section{Discussion}

We have observed the Seyfert galaxy WKK 6092 at different wavelengths.
The resulting data are summarized in Table~\ref{seytab}.

\begin{table}
 \caption{Observational parameters of WKK 6092}
 \label{seytab}
   \begin{tabbing}
   Apparent blue (IIIaJ)(m$_{B}$) magn  \=    \kill

{\bf Coordinates:} \> \\
   R.A. (1950)        \> $16^{h} 07^{m} 32.7^{s}$ \\
   DEC. (1950)        \> $-60^{\circ} 30' 11''$ \\
   Galactic Longitude \> $325.20^{\circ}$ \\
   Galactic Latitude  \> $-6.74^{\circ}$ \\

\vspace{5mm}
{\bf Properties:} \> \\
   Hubble Type        \> SBa(r) \\
   Dimensions (a x b) \> 56'' x 47'' \\
   Ellipticity (1-b/a)  \> 0.11 \\
   Inclination \> 28$\deg$ \\
   Position Angle \> 96$\deg$  \\

\vspace{5mm} 
{\bf Photometry:} \> \\
   $B_{\rm J}$ (IIIaJ)      \> 14.7 $\pm$ 0.5 mag \\
   $B_{25}$ (CCD)     \> 14.96 $\pm$ 0.09 mag \\
   $B_{\rm T}$ (CCD)      \> 14.88 $\pm$ 0.13 mag \\
   $R_{24}$ (CCD)     \> 13.38 $\pm$ 0.12 mag \\
   $R_{\rm T}$ (CCD)      \> 13.30 $\pm$ 0.14 mag \\
   J$_{\rm c}$              \> 12.91 $\pm$ 0.03 mag \\
   H                  \> 11.98 $\pm$ 0.03 mag \\
   K                  \> 11.51 $\pm$ 0.03 mag \\
   L                  \> 10.60 $\pm$ 0.20 mag \\
   \HI\ flux          \> 0.93 Jy \kms \\
   X-Ray (0.5--2.0 keV): \>    \\
   \hspace{0.25cm} Flux \> $1.05 \pm 0.15 \cdot 10^{-12}$ erg s$^{-1}$ \\
          \>   \hspace{2.65cm}       cm$^{-2}$ \\
   \hspace{0.25cm} ${\cal L}_X$ \> $1.2 \pm 0.17 \cdot 10^{42}$ 
                   erg s$^{-1}$\\

\vspace{5mm} 
{\bf  Galactic Extinction (A$_B$):} \>  \\
   from HI      \> 1.5 mag \\
   from Balmer decrement \> $\la$ 1.7 mag \\
   from X-ray   \> 1.6 mag \\

\vspace{5mm} 
{\bf Heliocentric velocity:}       \> \\
   MEFOS        \> 4711 $\pm$ 30  \kms \\
   S.A.A.O.     \> 4688 $\pm$ 40 \kms \\
   Parkes 64-m  \> 5012 $\pm$ 5 \kms \\
   \hspace{0.25cm} \DVF         \> 88 \kms \\
   \hspace{0.25cm} \DVT         \> 97 \kms \\

   \end{tabbing}
\end{table}

WKK 6092 and its neighbour are both members of ACO 3627. They have 
similar redshifts but show no indications of interaction. The 
morphology of both galaxies do not seem distorted (\cf\ Fig.~\ref{ccdim}). 
The Seyfert has a very blue nucleus, a distinct bar and a ring
superimposed on an otherwise smooth disk.

An upper limit for the Galactic foreground extinction in the line of sight of 
the Seyfert galaxy can be set at A$_{\rm B}$ = 1.6 mag. This was determined by three
different methods, the Balmer decrement in the optical spectrum, the fitting 
of an absorbed power low to the X-ray spectrum and the Galactic \HI\ column 
densities. All give a consistent value of the foreground extinction.  
A minor fraction of the extinction is intrinsic to the galaxy. 

The extinction corrected near-infrared colours of WKK 6092 are typical 
of a Seyfert 1 and are in agreement with well known Seyfert 1's such as 
NGC 1566 (Glass and Moorwood 1985). The X-ray sepctrum is also consistent 
with the standard expectation for this object. 

At the adopted cluster distance of R = 93 $h_{50}^{-1}$ Mpc, the
absolute magnitude (corrected for the galactic extincton) is M$_{B_T}^o = -21.52$.
The \HI\ and total mass is $1.9 \cdot 10^{9} {\cal M}_{\odot}$ and 
$30 \cdot 10^{9} {\cal M}_{\odot}$, respectively.
The Seyfert is at a projected distance of 0.8 $h_{50}^{-1}$ Mpc from 
the cluster centre and the \HI\ content of the galaxy might be
influenced by interactions with the Inter Cluster Medium due to 
processes like ram pressure stripping (Cayatte \etal\ 1990). The
\HI\ content is in fact quite low. This is, however,  
not inconsistent with the expectation for a barred early-type spiral.

Despite the difficulties in analysing data of an object deep within the Milky 
Way, all data concerning the here investigated Seyfert galaxy WKK 6092 at 
30 arcmin from the centre of the rich cluster ACO 3627 correspond to
the standard characteristics of a Seyfert 1 galaxy. 

\acknowledgements
PAW and APF are supported by the South African FRD. The research by RCKK has
been supported by an EC-grant.

\vfill
\end{document}